\documentstyle[preprint,eqsecnum,aps,epsfig]{revtex}
\tighten
\begin{document}
\draft
\preprint{UL--NTZ 11/98}
\title{
Production of relativistic positronium in collisions \\ of
photons and electrons with nuclei and atoms}
\author{S. R. Gevorkyan and E. A. Kuraev}
\address{Joint Institute of Nuclear Research, Dubna, 141980 Russia}
\author{A. Schiller}
\address{Institut f\"ur Theoretische Physik and NTZ, 
         Universit\"at Leipzig, \\ D-04109 Leipzig, Germany}
\author{V. G. Serbo}
\address{Novosibirsk State University, Novosibirsk, 630090 Russia}
\author{A. V. Tarasov}
\address{Joint Institute of Nuclear Research, Dubna, 141980 Russia}
\date{August 17, 1998}

\maketitle

\begin{abstract}
We consider the production of ultrarelativistic positronium
(Ps) in $\gamma  A  \to  {\mathrm {Ps}} + A $ and 
$e A \to {\mathrm{ Ps}} + e A$ processes where $A$ is an atom or a nucleus 
with charge $Ze$. For the photoproduction of para- and ortho-Ps and the
electroproduction of para-Ps we obtain the most complete description 
compared with previous works. It includes high order $Z\alpha$ corrections 
and polarization effects. The accuracy of the obtained cross sections is 
determined by  omitted terms of the order of the inverse Ps Lorentz factor. 
The studied high order multi-photon electroproduction of ortho-Ps 
dominates  for the collision of electrons with heavy atoms over the 
bremsstrahlung production from the electron via a virtual photon proposed by 
Holvik and Olsen. Our results complete and correct the studies of those authors.
\end{abstract}

 
\section{Introduction}

The production of relativistic positronium (Ps) opens an attractive possibility
to create intensive beams of elementary atoms. It is known that $e^+e^-$ 
elementary atoms exist in two spin states: parapositronium 
(para-Ps, singlet state) $^1\!S_0$ with lifetime at rest $\tau_0=0.123$ ns and 
orthopositronium (ortho-Ps, triplet state) $^3\!S_1$ with $\tau_0=0.14 \;\mu$s.
The relativistic Ps with lifetime $\tau = \gamma_{\mathrm{Ps}} \tau_0$ 
(where $\gamma_{\mathrm{Ps}}$ is its Lorentz factor) can be detected far from 
its creation point which is quite useful from the experimental point of view.

The main motivations to study the positronium production can be summarized as
follows: {\it (i)} It is the simplest hydrogen-like atom which is very 
convenient for testing fundamental properties of nature such as the CPT-theorem
(see review~\cite{M}); {\it (ii)} Up to now there is an essential difference 
between experimental measurements and theoretical calculations of the ortho-Ps 
width \cite{KM}; {\it (iii)} Finally, the relativistic Ps has an unusual large 
transparency in thin  layers (see Refs.~\cite{N} and literature therein), in 
QCD a similar property is called color transparency.

In this paper we consider both the Ps production in the collision of high 
energy photons and electrons with nuclei or atoms $A$ of charge $Z e$ 
(Figs.~\ref{Fig:1} and \ref{Fig:2}):
\begin{equation}
\gamma  A  \to  {\mathrm {Ps}}+A \,, \ \ \ 
eA  \to {\mathrm{Ps}}+ e A  \,.
\label{react} 
\end{equation}
Due to charge parity conservation, the number of exchanged photons between the
projectile photon and the target nucleus has to be odd or even for the 
production of para-Ps or ortho-Ps, respectively. 

Let us briefly summarize the present status of calculating the photo- and 
electroproduction of relativistic positronium. The photoproduction of para-Ps 
was calculated in main logarithmic approximation in \cite{MSS} and more 
precisely in \cite{O}. In both papers effects of high order corrections in the 
parameter $\nu$
\begin{equation}
\nu =Z\alpha = Z \frac{e^2}{4 \pi} \approx Z/137
\label{1}
\end{equation}
have not been taken into account. In (\ref{1}) $Ze$ denotes the nucleus charge.
However, the parameter $\nu$ is of the order of 1 for heavy nuclei and, 
therefore, the whole series in $\nu$ has to be summed. These $\nu$ effects were
considered for para- and ortho-Ps photoproduction in \cite{TC}, but only for 
total cross sections with complete screening. Polarization effects of the 
photon projectile and azimuthal asymmetries in the Ps distribution have not 
been studied. The electroproduction of para-Ps was calculated 
in~\cite{MSS,AB,HO} without high order $\nu$ corrections. 
The corresponding ortho-Ps was considered in \cite{AB,HO} where only the 
bremsstrahlung production of Fig.~\ref{Fig:3} was examined. 

The outline of our paper is as follows: In Sect.~\ref{sec:2} we obtain the most
complete description of the para- and ortho-Ps photoproduction on nuclei and 
atoms taking into account polarization, high order corrections in $\nu$ and
screening effects of target atoms. First the general matrix elements for the 
Ps production in virtual photon nucleus scattering is presented. Differential 
and total cross sections for para- and ortho-Ps photoproduction
are calculated in the following subsections. Sect.~\ref{sec:3} is devoted to 
the electroproduction of relativistic positronium. Our exact results for the 
para-Ps production are compared with calculations using the equivalent
photon approximation and neglecting high order $\nu$ contributions.
In Sect.~\ref{sec:33} we present a new mechanism for electroproduction of 
ortho-Ps, namely, the multi-photon production (summed over two, four, six, ... 
exchanged photons) of Fig.~\ref{Fig:2}. We show that it may be more important
than the bremsstrahlung mechanism of Fig.~\ref{Fig:3}, and even dominates for 
heavy atoms. In the final section we summarize our results.

Our main notations are collected in Figs.~\ref{Fig:1}, \ref{Fig:2}:
a photon with 4-momentum $q$ and energy $\omega$ or an electron with 4-momentum
$p_e$ and energy $E_e$ collides with a nucleus of 4-momentum $P$, mass $M$ and 
charge $Ze$ and produces Ps described by 4-momentum $p$, energy $E$ and 
polarization 4-vector $e$. We denote by $m_e$ the electron mass, the mass of 
the $e^+e^-$ bound state and a convenient abbreviation are given as follows 
\begin{equation}
m_{{\mathrm Ps}}\approx 2m_e,\; \;\;
\sigma_0 = \pi \nu^2 {\alpha^4\over m_e^2} \,.
\label{sigma0} 
\end{equation}
In the electroproduction of Ps (Fig.~\ref{Fig:2}) we deal with a virtual
photon generated by the electron. For this photon we use the notations
\begin{equation}
Q^2=-q^2 >0,\;\; y=\frac{Q^2}{4m_e^2},\;\; \mu = m_e\sqrt{1+y}.
\label{2}
\end{equation}
Throughout the paper we consider the production of relativistic Ps,
that means we consider the energy region
\begin{eqnarray}
\omega\approx E &\gg& 2m_e \ \ \ {\mathrm {for} }\ \ \ \gamma A \to
{\mathrm{Ps}}+ A\,, \nonumber \\ 
E_e > \omega\approx E &\gg& 2m_e \ \ \ {\mathrm {for}} \ \ \ eA \to 
{\mathrm{Ps}}+eA \,.
\label{3}
\end{eqnarray}

\section{Photoproduction of positronium}
\label{sec:2}

\subsection{
General matrix element for reaction $\gamma^* A \to {\mathrm{Ps}} + A$
}
\label{sec:21}

In this subsection we assume that the photon $q$ is virtual, $q^2 =-Q^2 <0$, 
and obtain formulas which will be useful both for photo- and electroproduction 
of Ps. Choosing the $z$ axis along the photon momentum, the polarization vector
for the transverse photon ($T$-photon $\gamma^*_T$ with helicity 
$\lambda_\gamma = \pm 1$) is
$$
e_{\gamma}=(0, e_{\gamma x},e_{\gamma y},0)\,.
$$
Taking into account gauge invariance, the polarization vector for the scalar 
photon ($S$-photon $\gamma^*_S$ with helicity $\lambda_\gamma = 0$) can be 
chosen in the form
$$
e_{S}= \frac{2\sqrt{Q^2}}{s}\,p_2\,,\;\;
p_2 =  P -\frac {M^2}  {s}\, q,\;\;\; s=2qP =2\omega M\,.
$$
In what follows, we calculate all distributions in an accuracy
neglecting only pieces of the order of
\begin{equation}
\frac { 2m_e }{\omega}\,, \;\;
\frac{|{\mathbf p}_\perp|}{\omega}\,, \;\;
\frac{\sqrt{Q^2}}{\omega}\,.
\label{18}
\end{equation}
Here ${\mathbf p}_\perp$ is the transverse component of the Ps 3-momentum.
In this accuracy the momentum transfer squared to the nucleus is given by
\begin{equation}
t=(p-q)^2=-4\mu^2(\tau+\epsilon^2)\,,\;\;
\tau = \frac {{\mathbf p}_\perp^2} {4 \mu^2}\,,\;\;
\epsilon=\frac{\mu}{\omega} \,.
\label{4}
\end{equation}

The amplitudes of the $\gamma^* A \to \mathrm{Ps} +A$ process are obviously 
closely related to those of the $e^+e^-$  pair production in the field of a 
heavy nucleus (for equal electron and positron momenta 
${\mathbf p}_+= {\mathbf p}_-={\mathbf p}/2 $). For the latter amplitudes we 
use the convenient form which was recently obtained 
in~\cite{IM}\footnote{Compare Eqs.~(29-30)
and (40-41) of~\cite{IM} taking into account the equal momenta for $e^\pm$. 
Note that in the paper of Ivanov and Melnikov 
an unusual notation for $\hat {\mathbf a}$ has been 
used what we would like to mention here:  
$\hat {\mathbf a} = \hat a$ where $\hat a = a_\mu \gamma^\mu$. 
In~\cite{IM} it has also been noticed that 
for the lepton 
pair production by a real incident photon, their expression for the amplitude 
(29) coincides with the known
result in the literature~\cite{BMOM,BLP}.
}:
\begin{equation}
M(\gamma^*_T A \to e^+e^- A) =-{Z(4\pi\alpha)^{3/2}\over \mu^3}
\bar{u}\hat{p}_2\left[({\mathbf n} {\mathbf e}_{\gamma} +
\hat{n} \hat{e}_{\gamma})\,\Phi_s+
{\mathrm i} \nu {2m_e \over \mu} \hat{e}_{\gamma} \,\Phi_t\right] v\,,
\label{5}
\end{equation}
\begin{equation}
M(\gamma^*_S A \to e^+e^- A) ={Z(4\pi\alpha)^{3/2}\over \mu^3}\,
{\mathrm i} \nu {\sqrt{Q^2} \over \mu} \,\Phi_t\,\bar{u}\hat{p}_2 v\,.
\label{6}
\end{equation}
Here $u$ and $v$ are the spinors of the produced pair of electron and positron,
furthermore the 4-unit vector $n$ is defined as 
$$
n= (0,\;{\mathbf n},\;0),\;\;\;
{\mathbf n} = \frac{{\mathbf p}_\perp}{|{\mathbf p}_\perp |} \,. 
$$
The functions $\Phi_s$ and $\Phi_t$ are given with the help of the Gauss 
hypergeometric function $F(a,b;c;z)$:
\begin{equation}
\Phi_s\equiv \Phi_s (\tau,\nu,\epsilon)={\sqrt{\tau}
\over (\tau+\epsilon^2) (1+\tau)}
F({\mathrm i} \nu,-{\mathrm i}\nu;1;z) \, {\pi\nu \over \sinh{\pi\nu}}\,,
\label{7}
\end{equation}
\begin{equation}
\Phi_t \equiv \Phi_t (\tau,\nu)= {1-\tau \over (1+\tau)^3}\,
F(1+{\mathrm i} \nu,1-{\mathrm i}\nu;2;z) \, {\pi\nu \over \sinh{\pi\nu}}\,,
\label{8}
\end{equation}
\begin{equation}
z= \left( {1-\tau \over 1+\tau} \right)^2 \,.
\end{equation}

Let us briefly mention some main properties of these functions:
\begin{eqnarray}
\Phi_s \to {\sqrt{\tau}\over (\tau+\epsilon^2) (1+\tau)}\,, \;\;
\Phi_t \to {1\over 1-\tau^2}\, \ln{{ (1+\tau)^2 \over 4\tau}}
\;\;\;  &{\mathrm at}& \;\; \nu \to 0\,,
\nonumber \\
\Phi_s \to {\sqrt{\tau}\over (\tau+\epsilon^2)}\,, \;\;
\Phi_t \to \ln{{ 1\over 4\tau}} -2 f(\nu)
\;\;\;&{\mathrm at}& \;\; \tau \to 0\,,
\nonumber \\
\Phi_s \to \frac{1}{2} 
\frac{\pi\nu}{\sinh{\pi\nu}}\,,
\;\;
\Phi_t \to \frac{1}{8} (1-\tau)\frac{ \pi \nu}{\sinh{\pi\nu}} 
\;\;\;&{\mathrm at}& \;\; \tau \to 1\,,
\label{Phiprop} \\
\Phi_s \to {1\over \tau^{3/2}}\,, \;\;
\Phi_t \to - {1\over \tau^2} \left[ \ln{{ \tau\over 4}} -2
f(\nu)\right] \;\;\;&{\mathrm at}& \;\; \tau \to \infty \,.
\nonumber
\end{eqnarray}
The function $f(\nu)$ is well known (see Eq. (95,19) in~\cite{BLP})
\begin{equation}
f(\nu) = \nu^2 \sum_{n=1}^{\infty} {1\over n(n^2+\nu^2)} \,.
\label{12}
\end{equation}
For small $\nu$ values it behaves as $f(\nu)\approx\zeta(3)\, \nu^2$
with the Riemann zeta function 
\begin{equation}
\zeta (3)= \sum_{n=1}^\infty {1\over n^3} = 1.2021 \,.
\end{equation}

As next step the  spinors of the electron and positron $\bar{u} \dots v$  
are substituted by the wave function of positronium in quantum state $n$ 
at the origin $\psi(0)=(m_e\alpha)^{3/2}/\sqrt{8\pi n^3}$  
according to the rule (see, for example,~\cite{NOSVVZ}):
\begin{equation}
\bar{u} \dots v \to {m_e\alpha^{3/2}\over \sqrt{4\pi n^3}}\;
\frac{1}{4}\; {\mathrm {Tr}} \; [\dots (\hat{p}+ 
m_{{\mathrm {Ps}}})\Gamma\,] \,.
\label{13}
\end{equation}
Here $\Gamma$ is a projection operator to be chosen  as
$\Gamma={\mathrm i} \gamma^5$ for the para-Ps and $\Gamma=\hat{e^*}$
for the ortho-Ps state.

Now the amplitudes of Ps production by $T$- or $S$-photons can be
obtained using (\ref{5}), (\ref{6}) and (\ref{13}). The para-Ps is produced only 
by transverse photons
\begin{equation}
M(\gamma^*_T A \to n ^1\!S_0+ A) ={ 2\pi Z \alpha^3\over
n^{3/2}} \, {s \over m_e^2 (1+y)^{3/2}}
\left( {\mathbf e}_{\gamma}\times {\mathbf n}\right)\cdot
{{\mathbf q}\over \omega}\;\Phi_s \,,
\label{14}
\end{equation}
\begin{equation}
M(\gamma^*_S A \to n ^1\!S_0+ A) = 0\,.
\label{15}
\end{equation}

For the ortho-Ps production there are two possibilities: The transversely  
polarized ortho-Ps (with helicity $\lambda = \pm 1$) arises from transverse 
photons only 
\begin{equation}
M(\gamma^*_T A \to n ^3\!S_1+ A) =-{\mathrm i} { 4\pi Z^2 \alpha^4\over
n^{3/2}} \, {s \over m_e^2 (1+y)^2}
( {\mathbf e}_{\gamma}\cdot {\mathbf e}^*)\;\Phi_t \,,
\label{16}
\end{equation}
while the longitudinally polarized ortho-Ps (with helicity $\lambda = 0$) 
is produced by scalar photons
\begin{equation}
M(\gamma^*_S A \to n ^3\!S_1+ A) 
= {\mathrm i}{\sqrt{Q^2} \over 2m_e}\,  { 4\pi
Z^2 \alpha^4\over n^{3/2}} \, {s \over m_e^2 (1+y)^2} \,\Phi_t \,.
\label{17}
\end{equation}
From (\ref{16}) and (\ref{17}) we observe that helicity is conserved in the
$\gamma \to$ ortho-Ps transition, i.~e. $\lambda_\gamma = \lambda$. 
We also notice that both amplitudes (\ref{16}), (\ref{17}) do not depend on the 
azimuthal angle of ortho-Ps.

In the following parts of this section we study the production of
positronium by real photons. Therefore, in all expressions we
have to choose (see (\ref{2})) $Q^2=0$, $y=0$.

\subsection{Photoproduction of para-Ps}
\label{sec:22}
The differential cross section of para-Ps production on nuclei can be
obtained using the amplitude (\ref{14}) at $Q^2=0$
\begin{equation}
d\sigma_{{\mathrm {singlet}}}={\sigma_0\over n^3}\,
|{\mathbf e}_{\gamma} \times {\mathbf n}|^2 \,\Phi_s^2\,
d\tau \,\frac{d\varphi}{2\pi}
\label{dssinglet}
\end{equation}
where $\varphi$ is the azimuthal angle of Ps. To describe the polarization
degree of the initial photon in general, it is convenient to use Stokes
parameters $\xi_{1,2,3}$. In that case (\ref{dssinglet}) transforms to
\begin{equation}
d\sigma_{{\mathrm {singlet}}}={\sigma_0 \over 2n^3} \,
 [1+\xi_1 \sin{2\varphi}- \xi_3 \cos{2\varphi}] \,\Phi^2_s\,
d\tau\, \frac{d\varphi}{2\pi}\,.
\label{19}
\end{equation}
We see that this cross section does not depend on $\xi_2$, i.e. on
the circular polarization of photon. Furthermore, the scattering
plane is mainly orthogonal to the direction of the linear
polarization of the photon.
For $\xi_i \to 0$ and $\nu \to 0$ the result (\ref{19})
coincides with that obtained in \cite{O}.

The dependence of cross section (\ref{19}) on the Ps polar angle
$\theta$ is completely described by the function $\Phi_s^2(\tau,
\nu, \epsilon)$ since in our case
\begin{equation}
\tau= \left({\omega\theta\over 2m_e} \right)^2\,.
\label{tau}
\end{equation}
According to (\ref{Phiprop}) the function
$\Phi_s(\tau,\nu,\epsilon)$ depends on $\nu$
only in the region $\tau\sim 1$, whereas for  $\tau \ll 1$
and $\tau \gg 1$ the angular behavior of cross section
(\ref{19}) has an universal (independent on $\nu$) character. In
particular, in the region of very small angles
\begin{equation}
\frac{m_e^2}{\omega^2} \ll \theta \ll \frac{m_e}{\omega}
\label{20}
\end{equation}
the differential cross section has a very simple form
$$
d\sigma_{{\mathrm {singlet}}}={\sigma_0 \over n^3} \,
{d\theta \over \theta}\,.
$$

The total cross section is obtained from (\ref{19}) by integrating
over $\varphi$ and $\tau$ and by summing over all
$n \, ^1\!S_0$ quantum states:
\begin{equation}
\sigma_{{\mathrm {singlet}}}=\sigma_0\,\zeta (3)
\, \left[\ln{{\omega\over m_e}}-1-C(\nu)\right]\,,
\label{21}
\end{equation}
\begin{eqnarray}
C(\nu)=\frac{1}{2}\int\limits_0^{\infty} \left\{ 1-\left[
\,F({\mathrm i}\nu,-{\mathrm i}\nu;1;z)\,
\frac{\pi\nu}{\sinh{\pi\nu}} \,\right]^2
 \right\} \frac{d\tau}{\tau(1+\tau)^2} \nonumber \\
=\frac{1}{4}\int\limits_0^1 \left\{1-\left[
\, F({\mathrm i}\nu,-{\mathrm i}\nu;1;z)\frac{\pi \nu}{\sinh{\pi\nu}}
\,\right]^2\right\} \frac{(1+z) dz}{(1-z)\sqrt{z}}.
\label{22}
\end{eqnarray}
The function $C(\nu)$ is presented in Fig.~\ref{fig4_c},
at small $\nu$ it is approximated by
\begin{equation}
C(\nu)=\left[ \frac{7}{2}\zeta (3)-4+4\ln 2\right] \nu^2\approx
2.9798\, \nu^2\,.
\label{23}
\end{equation}
Note that the large logarithmic term $\ln (\omega /m_e)$ in cross section 
(\ref{21}) arises just from the region of small angles (\ref{20}).
Therefore, this region  determines the characteristic polar angle of para-Ps 
production $\theta_{\mathrm{char}}^{^1\!S_0}$.

Up to now we have considered the photoproduction of positronium 
on nuclei. Let us briefly discuss the photoproduction  on atoms where
a possible atomic screening has to be taken into account. 
This can be done by inserting a factor $\left(1-F(t)\right)$
in the amplitude (\ref{14}) (and $\left( 1-F(t)\right)^2$ in the cross sections 
(\ref{dssinglet}) or (\ref{19})) where $F(t)$ is the atomic form
factor and $t$ is given in (\ref{4}). As a result, we obtain
the total cross section for photoproduction of para-Ps on atoms:
\begin{equation}
\sigma_{{\mathrm {singlet}}}= \sigma_0\, \zeta (3) \,
\left[J-C(\nu)\right]\,,\;\;\;
J=\frac{1}{2}\int\limits_0^{\infty}
\frac{\tau [1-F(t)]^2}{(\tau+\epsilon^2)^2(1+\tau)^2}\, d\tau \,.
\label{26}
\end{equation}

If we use a simplified Thomas-Fermi-Moli\'{e}re form factor
\begin{equation}
F(t)={1\over 1- (t/{\Lambda^2)}}\,,\;\;\;
\Lambda=\frac{Z^{1/3}}{111} \, m_e  \,,
\label{27}
\end{equation}
we obtain in accuracy (\ref{18})
\begin{equation}
J = - \frac{1}{2} \, \ln  \left[ \left( \frac{m_e}{\omega} \right)^2 +
\left( \frac {Z^{1/3}} {222} \right)^2  \right] -  1\,.
\label{28}
\end{equation}
At not very high energies
\begin{equation}
2m_e \ll \omega \ll \frac{2m_e^2}{\Lambda} =
\frac{222}{Z^{1/3}}\,m_e
\label{29}
\end{equation}
the screening effects are negligible in $J$ and the previous result for the 
nucleus target (\ref{21}) remains valid. For high enough energies
\begin{equation}
\omega \gg \frac{222}{Z^{1/3}}\,m_e\,,
\label{30}
\end{equation}
there is complete screening and the total cross section takes the form
\begin{equation}
\sigma_{{\mathrm {single}t}}= \sigma_0 \,\zeta (3) \,
\left[\ln{{222\over Z^{1/3}}}-1-C(\nu)\right]
\label{31}
\end{equation}
which coincides with the result obtained in~\cite{TC}.

\subsection{Photoproduction of ortho-Ps}
\label{sec:23}

The differential cross section of ortho-Ps production on nuclei can be 
obtained using the amplitude (\ref{16}). For polarized photons it is given by
\begin{equation}
d\sigma_{{\mathrm {triplet}}}=4\nu^2 \,\frac{\sigma_0}{n^3}\,
|{\mathbf e}_{\gamma}\cdot {\mathbf e}^*|^2\, \Phi_t^2 \, d\tau \,,
\label{32}
\end{equation}
for unpolarized photons we have
\begin{equation}
d\sigma_{{\mathrm {triplet}}}=4\nu^2 \,\frac{\sigma_0}{n^3}\,
\Phi_t^2 \,d\tau\,.
\label{33}
\end{equation}
The dependence of this cross section on $\tau$ (and therefore on the polar 
angle of positronium $\theta$, see (\ref{tau})) is given by the function 
$\Phi_t^2$, it is presented in Fig.~\ref{Fig:5}. 
$d \sigma_{\mathrm {triplet}} / d \tau$ vanishes for all values
of $\nu$ as $(1-\tau)^2$ at $\tau \to 1$ (see (\ref{Phiprop})).

Comparing (\ref{19}) and (\ref{33}) we conclude
that the angular distributions of  ortho-Ps production is
considerably wider than that of  para-Ps production. Indeed, the
typical value of $\tau$ for ortho-Ps production is of the
order of $0.1$ which corresponds to a characteristic emission angle
\begin{equation}
\theta_{{\mathrm char}}^{^3\!S_1} \sim {m_e\over \omega}\,,
\label{34}
\end{equation}
while for para-Ps production on nuclei the region of very small angles 
(\ref{20}) gives the main contribution to the cross section.

The total cross section, obtained from (\ref{33}), is independent on the 
energy of the initial photon:
\begin{equation}
\sigma_{{\mathrm {triplet}}}= 4 (Z\alpha)^2 \,\sigma_0\,\zeta (3) 
B(\nu)\,.
\label{35}
\end{equation}
Here the function $B(\nu)$ is
\begin{equation}
B(\nu)=\int\limits_0^\infty \Phi_t^2(\tau,\nu)d\tau=
\left(\frac{\pi\nu}{\sinh{\pi\nu}} \right)^2
\frac{1}{8}\int\limits_0^1 \,\sqrt{z}(1+z)\,
\left[\,F(1+{\mathrm i}\nu,1-{\mathrm i}\nu;2;z) \right]^2 \,dz\,,
\label{36}
\end{equation}
its dependence on $Z$ is shown in Fig.~\ref{fig6_b}. 
At small $\nu \ll 1$  this function behaves as \cite{TC}:
\begin{equation}
B(\nu)=2-2\ln 2\,- \,\left [8(2-\ln{2})^2-\frac{2}{3}\pi^2-5\zeta
(3) \right]\,\nu^2 \approx 0.6137- 1.0729 \nu^2\,.
\label{37}
\end{equation}

The obtained results for the photoproduction of ortho-Ps on nuclei are also 
valid for the production on atoms. Indeed, the typical value of 
$\tau^{^3\!S_1} \sim 0.1$ for photoproduction on nuclei is much
larger than the characteristic value
$$
\tau_{\mathrm screen} \sim \left({\Lambda\over 2m_e}\right)^2  = 
 \left( \frac {Z^{1/3}} {222} \right)^2
$$
for which we should take into account the atomic form factor (see (\ref{27})).

At the end of this section let us compare the photoproduction of
ortho- and para-Ps on atoms. For energies $\omega \gg 222 m_e/ Z^{1/3}$ 
the ratio
\begin{equation}
\frac{\sigma_{{\mathrm triplet}}} {\sigma_{{\mathrm singlet}}}=
\frac {4\nu^2 B(\nu)}{\ln{(222/Z^{1/3})} -1 - C(\nu)}
\label{38}
\end{equation}
is presented as function of the nucleus charge number $Z$  
in Fig.~\ref{Fig:ratios3s1}. Some particularly interesting values are
28.5~\%, 23.5~\% and 1.51~\% for U, Pb and Ca, respectively.

\section{Electroproduction of positronium}
\label{sec:3}
\subsection{General expressions for reaction $eA \to {\mathrm{Ps}} + eA$}
\label{sec:31}
It is well known that the cross section for the electroproduction
of Fig.~\ref{Fig:2} can be exactly written in terms of two structure
functions or two cross sections $\sigma_T(\omega, Q^2)$ and 
$\sigma_S(\omega, Q^2)$ for the processes
$\gamma_T^* A \to {\mathrm{ Ps}}+A$ and $\gamma_S^*A \to {\mathrm{Ps}}+A$: 
\begin{equation}
d\sigma(eA  \to {\mathrm {Ps}}+eA) = \sigma_T(\omega, Q^2)\,
dn_T(\omega, Q^2)
+ \sigma_S(\omega, Q^2)\,dn_S(\omega, Q^2) \,.
\label{39}
\end{equation}
Here the coefficients $dn_T$ and $dn_S$ can be called the number of
transverse and scalar virtual photons, respectively. 
Using the amplitudes (\ref{14})-(\ref{17}) for the corresponding processes 
the cross sections $\sigma_T$ and $\sigma_S$ are calculable for virtual photon
energies squared
\begin{equation}
\omega^2 \gg (2m_e)^2, \,Q^2
\label{40}
\end{equation}
with accuracy (\ref{18}). In the same region and with 
the accuracy $\sim (2m_e/\omega)^2, 
\; Q^2/ \omega^2$ the 
quantities $dn_T$ and $dn_S$ are (see, for example, Sect. 6 in review 
\cite{BGMS})
\begin{equation}
dn_T = {\alpha \over \pi}\, N\left({\omega \over E_e}, {Q^2 \over
4m_e^2}\right)\, {d\omega \over \omega}\, {dQ^2 \over Q^2}\,,\;\;
N(x,y) = 1-x+\frac{1}{2}x^2 -{x^2\over 4y}\,,
\label{41}
\end{equation}
\begin{equation}
dn_S = {\alpha \over \pi}\, \left(1-{\omega \over E_e}\right)\,
{d\omega \over \omega}\, {dQ^2 \over Q^2}\,.
\label{42}
\end{equation}
The variable $y$ is defined in (\ref{2}).
Since the energy $E$ of the positronium almost coincides with that of the
virtual photon, the energy fraction transferred from the electron to Ps is
\begin{equation}
x={E \over E_e} = {\omega \over E_e}.
\label{43}
\end{equation}

\subsection{Electroproduction of para-Ps}
\label{sec:32}
The cross sections $\sigma_T$ and $\sigma_S$ for  para-Ps production are 
obtained using (\ref{14}), (\ref{15}) and repeating the calculations of 
Sect.~\ref{sec:22} with $Q^2 > 0$. This leads to
\begin{equation}
d\sigma_T={\sigma_0 \over 2n^3 (1+y)^2} \,
\,\Phi^2_s\, d\tau\,, \;\; d\sigma_S=0\,,
\label{44}
\end{equation}
(with $\tau$ defined in (\ref{4})) and to the integrated  cross sections
\begin{equation}
\sigma_T={\sigma_0 \over (1+y)^2} \, \zeta(3) \, [L-1-C(\nu)]
\,, \;\;\sigma_S=0 \,.
\label{45}
\end{equation}
Here for the electroproduction on nuclei (compare  (\ref{19}), (\ref{21}))
\begin{equation}
L=\ln{{\omega \over m_e}} - \frac{1}{2} \ln{(1+y)}
\label{46}
\end{equation}
and on atoms (compare  (\ref{26}), (\ref{28}))
\begin{equation}
L=- \frac{1}{2}\ln{\left[\left({m_e \over \omega}\right)^2 (1+y) +
\left({\Lambda \over 2m_e}\right)^2 {1\over (1+y)} \right]}\,.
\label{47}
\end{equation}

Using (\ref{39})-(\ref{47}) we are able to obtain the energy-angular and energy 
distributions of relativistic Ps. In particular, the spectrum of para-Ps is
\begin{eqnarray}
&&d\sigma(eA \to ^1\!\!S_0 +eA) = {\alpha \over \pi} \,
\sigma_0 \, \zeta(3) F_s(x)\, {dx \over x}\,,
\label{48} \\
F_s(x) &=& \int\limits_{y_m}^\infty \,{[L-1-C(\nu)] \over (1+y)^2}\,
N(x,y) \,{dy\over y}\,, \;\;\; y_m = {x^2\over 4(1-x)}
\nonumber
\end{eqnarray}
(Since the cross section rapidly decreases  above $y\approx 1$,
the upper integration limit can be extended  to infinity.).

For the electroproduction on nuclei the $y$ integration is easily performed 
and we obtain
\begin{equation}
F_s(x)= f_1(x) \left[\ln{xE_e \over m_e}-1-C(\nu)\right] -f_2(x) \,.
\label{49}
\end{equation}
The functions $f_1(x)$ and $f_2(x)$ can be presented as follows:
\begin{eqnarray}
&& f_1(x)=2 (1-x+x^2) \,\ln{2-x\over x} -{4(1-x)\over
(2-x)^2}(2-2x+x^2)\,,
\nonumber \\
&& f_2(x)  = (1-x+x^2) \left[ \frac{\pi^2}{12} - \frac{2
 (1-x)}{(2-x)^2} -  \frac12 {\mathrm{ Li}} _2\left(
   \frac{x^2}{(2-x)^2}\right)  \right]
\label{f2xnew} \\
&& +  \frac{x^2}{4} \left( \frac{2 (1-x)}{(2-x)^2} + \ln
\frac{x}{2-x} \right) - \frac{2 (1-x)}{(2-x)^2} (2- 2 x + x^2)
   \ln \frac { (2-x)^2}{4 (1-x)}
\nonumber
\end{eqnarray}
with the dilogarithm function
$$
{\mathrm{ Li}}_2 (z)= \int\limits_{z}^{0} \frac{\ln (1-t)}{t}
dt\,.
$$
Note that $f_2(x) < f_2(0)=(\pi^2-6)/12=0.3225$.

The same result (\ref{49}) is valid for electroproduction on atoms at 
$2m_e \ll xE_e \ll 222m_e/Z^{1/3}$ (no screening). For the case of complete 
screening ($xE_e \gg 222m_e/Z^{1/3}$) we have
\begin{equation}
F_s(x)= f_1(x) \left[\ln{222 \over Z^{1/3}}-1-C(\nu)\right]
+f_2(x)\,.
\label{50}
\end{equation}
In the important case of small $x$ (i.e. at $2m_e/E_e \ll x \ll 1$)
the spectrum is simplified:
\begin{equation}
F_s(x)= \left( 2\ln{2\over x} -2 \right)\,
\left[\ln{xE_e \over m_e}-1-C(\nu)\right] -{\pi^2-6\over 12}
\label{51}
\end{equation}
for no screening and
\begin{equation}
F_s(x)= \left( 2\ln{2\over x} -2 \right)\,
\left[\ln{222 \over Z^{1/3}}-1-C(\nu)\right] + {\pi^2-6\over 12}
\label{52}
\end{equation}
for complete screening. The accuracy of the obtained spectrum is determined by 
omitting terms of the order of
\begin{equation}
\frac{2m_e} {xE_e} \,.
\label{53}
\end{equation}

It is interesting to compare our spectra (\ref{48})-(\ref{52}) with that given  in \cite{HO}
\begin{equation}
F_s^{{\mathrm {HO}}}(x) = \left( 2\ln{1\over x}\,-\,1 \right)
\left\{- \frac{1}{2}\ln{\left[\left({m_e \over xE_e}\right)^2
+ \left({Z^{1/3} \over 222}\right)^2 \right]} \,-\,1\right\}\,.
\label{54}
\end{equation}
The spectrum $F_s^{{\mathrm {HO}}}(x)$ was obtained neglecting the exact 
dependence of $\sigma_T$ on $Q^2$ and  high order $\nu$ corrections.
As a consequence, the accuracy of (\ref{54}) is only
logarithmic (in expressions (\ref{51})-(\ref{52}) and (\ref{54})
the leading logarithmic terms coincide whereas the next to leading logarithmic 
terms are different even at $\nu \ll 1$). Therefore, this approximation is not 
well suited for heavy atoms. For example, in the case of complete screening the
spectrum $F_s^{{\mathrm {HO}}}(x)$ exceeds our spectrum $F_s(x)$ up to 
approximately 40~\% for U and 30~\% for Pb in the $x$ region below 0.4.

\subsection{Electroproduction of ortho-Ps}
\label{sec:33}
The ortho-Ps production in collisions of electrons with atoms due to the 
bremsstrahlung mechanism of Fig.~\ref{Fig:3} was calculated in \cite{AB,HO}. 
The principal features of this mechanism are the following: The spectrum of 
ortho-Ps has a peak in the region of high energy fractions (at $x \approx 1$), 
the characteristic emission angle of Ps is small
\begin{equation}
\theta^{\mathrm {br}}_{\mathrm {char}}\sim\frac{m_e}{E_e} \,.
\label{thetacharbr}
\end{equation}
The total cross section is equal to~\cite{HO}
\begin{equation}
\sigma_{\mathrm {br}}={\alpha\over \pi}\, \sigma_0\, \zeta(3)\,
I_{\mathrm {br}}
\label{55}
\end{equation}
where
\begin{equation}
I_{\mathrm {br}}=0.303\ln{E_e\over m_e}\,-0.542
\end{equation}
for no screening ($E_e \ll 444 m_e /Z^{1/3}$) and
\begin{equation}
I_{\mathrm {br}}=0.303\ln{111\over Z^{1/3}}\,+0.362
\label{Ibr2}
\end{equation}
for complete screening ($E_e \gg 444 m_e /Z^{1/3}$).

In this section we argue that in many respects these results are incomplete or 
even misleading because in~\cite{AB,HO} the important multi-photon production 
(MP) of ortho-Ps due to diagrams of Fig.~\ref{Fig:2} with even numbers 
$j=2,4,6,\dots$ of exchanged photons was not considered. Moreover, we find out 
that MP production is dominant for electron scattering on heavy atoms.

To study the MP production of ortho-Ps, we have to calculate the cross 
sections $\sigma_T$ and $\sigma_S$ for ortho-Ps production (see (\ref{39})). 
This is achieved using the corresponding amplitudes (\ref{16}), (\ref{17})  
and repeating the  calculations of Sect.~\ref{sec:23} with $Q^2 > 0$. For the 
cross sections we obtain (compare (\ref{33}), (\ref{35}))
\begin{equation}
d\sigma_T= 4\nu^2 \,\frac{\sigma_0}{n^3(1+y)^3}\,
\Phi_t^2 \,d\tau\,,\;\;
d\sigma_S= y\, d\sigma_T\,,
\label{56}
\end{equation}
\begin{equation}
\sigma_T= 4 \nu^2 \,{\sigma_0 \over (1+y)^3}\,\zeta (3)\,
B(\nu)\,,\;\;\;
\sigma_S= y\, \sigma_T\,.
\label{57}
\end{equation}

Using these formulas and (\ref{39}) we can again obtain the energy-angular and 
energy distributions for the triplet state of relativistic positronium. In
particular, now the spectrum of ortho-Ps is
\begin{equation}
d\sigma_{{\mathrm MP}} = {4\alpha \over \pi} \,(Z\alpha)^2\, \sigma_0
\, \zeta(3) B(\nu) \,F_t(x)\, {dx \over x}\,,
\label{58} 
\end{equation}
\begin{equation}
F_t(x) =2 \left(1-x+\frac{5}{4} x^2\right)\, \ln{2-x\over x}
-{(1-x)\over (2-x)^4}\,[32(1-x)^2+34(1-x)x^2+5x^4]\,.
\label{58a}
\end{equation}
The spectrum (function $F_t(x)/x$) is shown in Fig.~\ref{fig7}. The result 
(\ref{58}), (\ref{58a}) is valid for collisions of electrons both with nuclei 
and atoms. The behavior of function $F_t(x)$ at small and large $x$ is the 
following:
\begin{equation}
F_t(x) = 2\ln{2\over x} -2 \;\;\; {\mathrm {at} } \;\; x\ll 1\,, \;
F_t(x) = {56\over 3} (1-x)^3 \;\;\; {\mathrm {at} } \;\;1- x\ll
1\,.
\label{59}
\end{equation}
Contrary to the bremsstrahlung spectrum (see Fig.~\ref{Fig:3} in \cite{HO}), 
the spectrum (\ref{58}) is peaked at small energies of Ps. Therefore, the 
interference of bremsstrahlung and MP production should be very small. 

Taking into account (\ref{34}) and (\ref{43}), we conclude that the 
characteristic angle for MP production of ortho-Ps is
\begin{equation}
\theta^{{\mathrm {MP}}}_{\mathrm {char}}\sim \frac{m_e}{xE_e}
\label{60}
\end{equation}
which is much larger than (\ref{thetacharbr}). In other words, the angular 
distribution of MP production is
considerably wider than that of the bremsstrahlung reaction. 
 
In this section the spectra  for relativistic positronium in electroproduction 
have been calculated with the high accuracy (\ref{53}). This accuracy cannot be
achieved for the total cross sections in the scheme used here by the following 
reasons: The spectrum has to be integrated over the whole kinematic region in 
$x$ including that of the threshold $x \sim 2m_e/E_e$. But near the threshold 
the accuracy of our calculated spectra becomes only logarithmic. Furthermore, 
in this region Ps is not a relativistic particle and, therefore, its detection 
is difficult. However, if we are interested to find the total cross section 
with logarithmic accuracy, we can integrate the spectra in the whole region
\begin{equation}
x_m = {2m_e\over E_e} \le x \le 1\,.
\label{61}
\end{equation}

Having in mind this restriction, we find for the total cross section of 
MP production 
\begin{eqnarray}
&& \sigma_{{\sc MP}} \approx {4\alpha \over \pi} \,(Z\alpha)^2\,
\sigma_0 \, \zeta(3) B(\nu) \, I_{\sc MP}\,,
\label{62} \\
 I_{\mathrm {MP}}= \int\limits_{x_m}^1\,F_t(x)\, {dx \over x}
&\approx& \left(\ln\frac{E_e}{m_e}\right)^2 - 2 \ln\frac{E_e}{m_e}-c\,,
\; \;
c={\pi^2\over 6} -{5\over 4} =0.3949 \,.
\nonumber
\end{eqnarray}
Note that this multi-photon cross section increases with the energy of the 
projectile electron as $[\ln{(E_e/m_e)}]^2$ while the cross section for the 
bremsstrahlung production on atoms is constant at high energies 
(see (\ref{Ibr2})). The dependence of the cross section ratio
\begin{equation}
\frac {\sigma_{\mathrm {MP}}}  {\sigma_{\mathrm {br}}}=
4\nu^2\,B(\nu)\frac{I_{\mathrm {MP}}} {I_{\mathrm {br}}}
\label{63}
\end{equation}
on the initial electron energy $E_e$ is presented in Fig.~\ref{fig8}
for different values of $Z$. From that picture it can be seen 
that the MP production is the dominant production mechanism of orthopositronium
for atoms with nucleus charge number $Z> 20$.

\section {Summary}
\label{sec:4}

In this paper we have presented an almost complete description of the 
production of relativistic positronium in high energy photon and electron 
collisions with nuclei and atoms. The high accuracy of our results is restricted
by neglecting terms of the order of the inverse Ps Lorentz factor  
$2 m_e/ E$.

The matrix elements of the virtual photon nucleus scattering to produce both 
para- and ortho-Ps are given in (\ref{14})-(\ref{17}) including 
polarizations of the initial photon and the positronium and summed
high order $Z \alpha$ corrections. The singlet positronium state can be 
produced only by transversely polarized initial photons, the transition 
amplitude depends on the azimuthal angle of $^1\!S_0$. The amplitude for scalar
virtual photons to para-Ps is zero. The transitions from the initial virtual 
transverse and scalar photon to the triplet state are accompanied with helicity
conservation $\lambda_{\gamma}=\lambda$, the amplitudes do not depend on the 
azimuthal angle of ortho-Ps.

These results are used to discuss both photo- and electroproduction of Ps.
Various distributions and total cross sections are calculated and compared to 
previous results. The screening effects are estimated  analytically
using a Thomas-Fermi-Moli\'{e}re atomic form factor.

For the photoproduction of relativistic positronium we have found that the 
polar angular distribution of ortho-Ps is considerably wider than that of 
para-Ps. The high order $Z \alpha$ effects decreases the ortho-Ps 
photoproduction cross section by 3.61~\% for Ca, 40.5~\% for Pb and 46.5~\% for
U nuclei. The ratio of the total cross sections for the triplet to singlet 
state at higher energies $\omega\gg 222 m_e/Z^{1/3}$ in the 
$\gamma A \to ^3\!\!S_1 + A$ process raises with the nucleus charge number from
1.51~\% for Ca, 23.5~\% for Pb to 28.5~\% for U targets.

In the para-Ps electroproduction the virtuality of the photons arising from the
electron projectile and the effects of heavy nuclei are quite important. As an 
example, the spectrum \cite{HO} estimated in an equivalent photon approximation
and with neglected high order $Z \alpha$ effects exceeds the correctly 
calculated result up to 30~\% for Pb and 40~\% for U in a wide range of the 
energy fraction transferred from the electron to Ps. 

Finally we have proposed a new multi-photon mechanism for the production of 
ortho-Ps in the reaction $e A \to ^3\!\!S_1 + e A$ which has to be taken into 
account besides the bremsstrahlung production discussed by Holvik and Olsen.
Due to a completely different angular and energy distribution of MP its
interference with the bremsstrahlung reaction is expected to be small. This new
mechanism is dominant for electron scattering on heavy atoms. Therefore, our 
results complete and correct those earlier studies.

\begin{figure}[!htb]
  \centering
  \epsfig{file=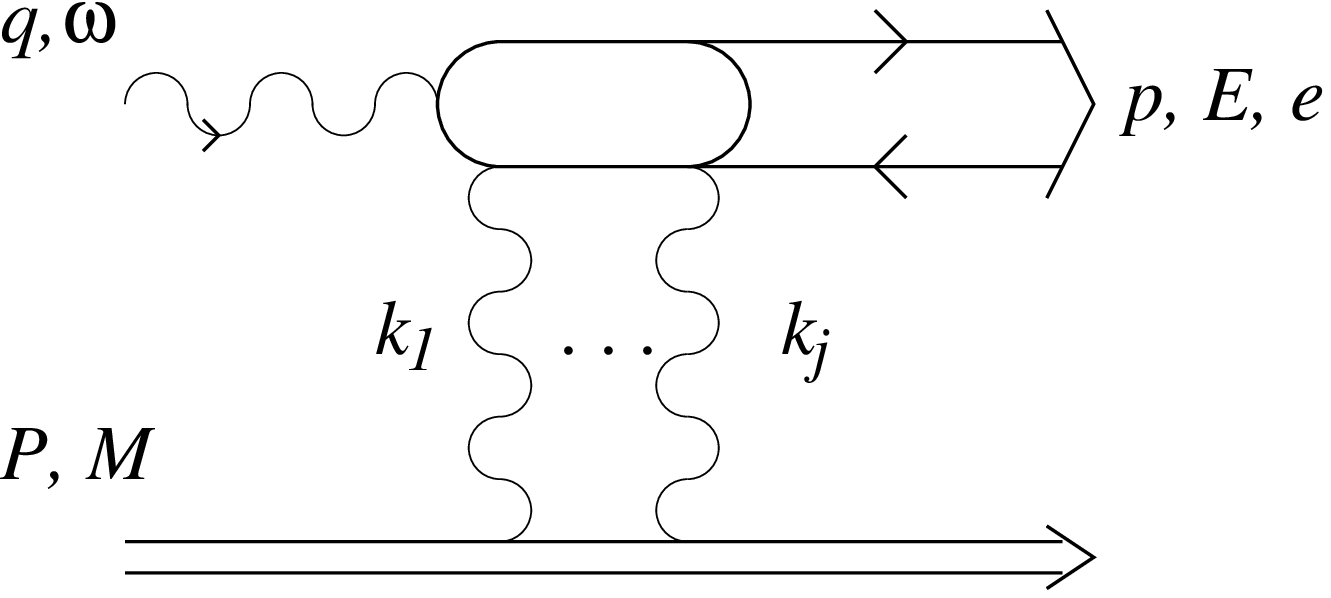,height=70mm,angle=270}
  \vspace{3mm}
 \caption{Photoproduction of Ps on nucleus with odd (even)
 number $j$ of exchanged photons   
 for para-Ps (ortho-Ps).}
\label{Fig:1}
\end{figure}

\begin{figure}[!htb]
  \centering
  \epsfig{file=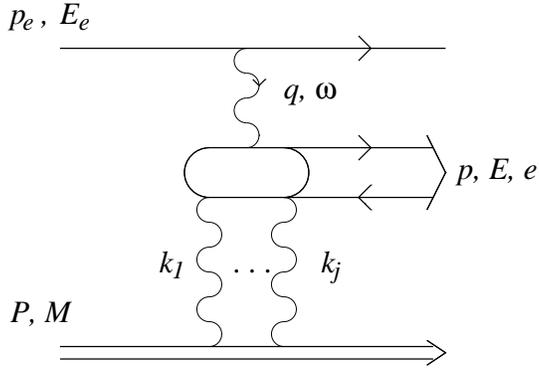,height=70mm,angle=270}
  \vspace{3mm}
 \caption{Electroproduction of Ps on nucleus.}
   \label{Fig:2}
\end{figure}

\begin{figure}[!htb]
  \centering
  \epsfig{file=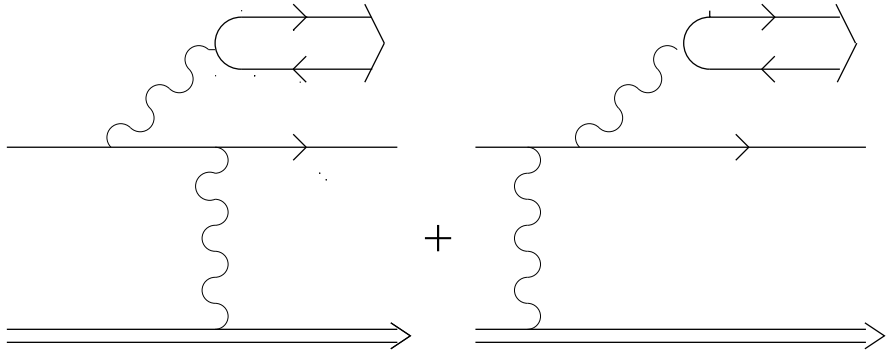,height=120mm,angle=270}
  \vspace{3mm}
 \caption{Bremsstrahlung production of ortho-Ps on nucleus.}
   \label{Fig:3}
\end{figure}

\begin{figure}[!htb]
  \centering
  \epsfig{file=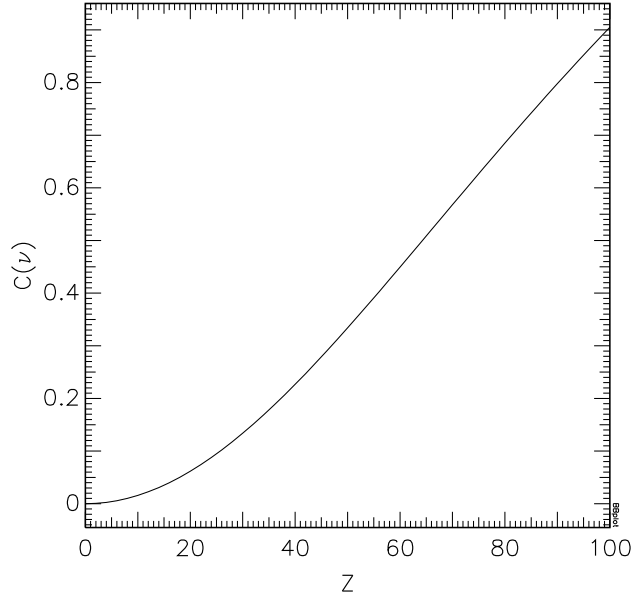,height=80mm}
  \caption{Function $C(\nu)$ (\ref{22}) vs. nucleus charge number $Z$.}
  \label{fig4_c}
\end{figure}

\begin{figure}[!htb]
  \centering
  \epsfig{file=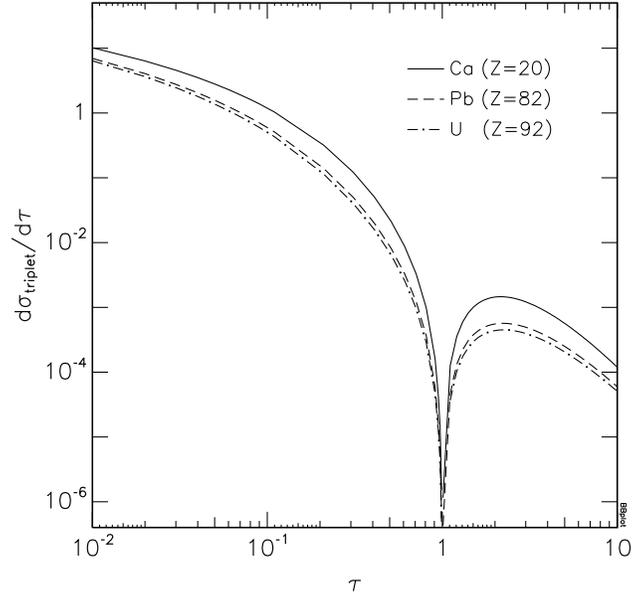,height=80mm}
  \caption{Differential cross section $d\sigma_{\mathrm {triplet}}/
  d\tau$ in units $4\nu^2\sigma_0/n^3$.}
\label{Fig:5}
\end{figure}

\begin{figure}[!htb]
  \centering
  \epsfig{file=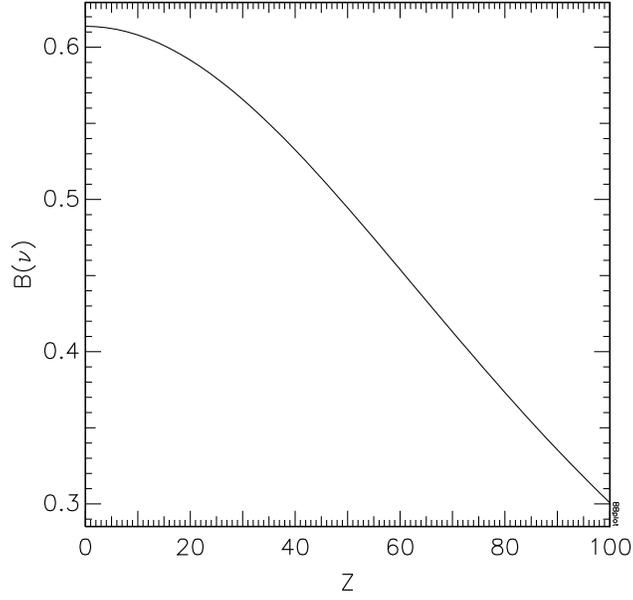,height=80mm}
  \caption{Function $B(\nu)$ ((\ref{36})) vs. $Z$.}
  \label{fig6_b}
\end{figure}

\begin{figure}[!htb]
  \centering
  \epsfig{file=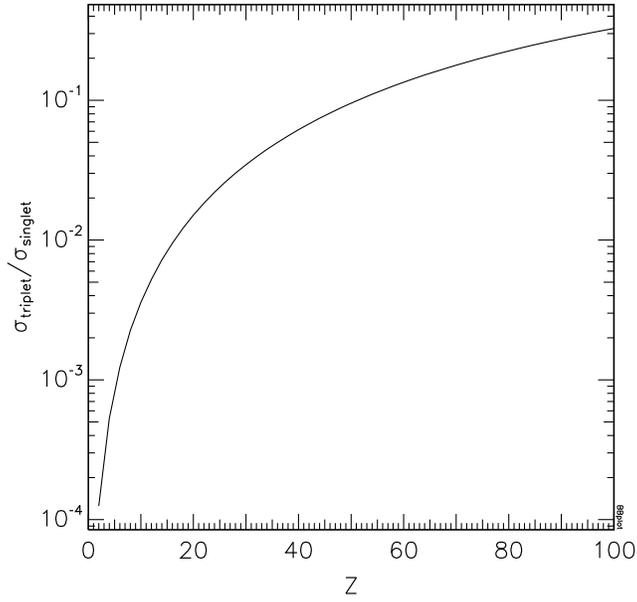,height=80mm}
  \caption{Ratio $ \sigma_{\mathrm {triplet}}/ \sigma_{\mathrm {singlet}}$
  in photoproduction on atoms at larger energies as function of $Z$.}
  \label{Fig:ratios3s1}
\end{figure}

\begin{figure}[!htb]
  \centering
  \epsfig{file=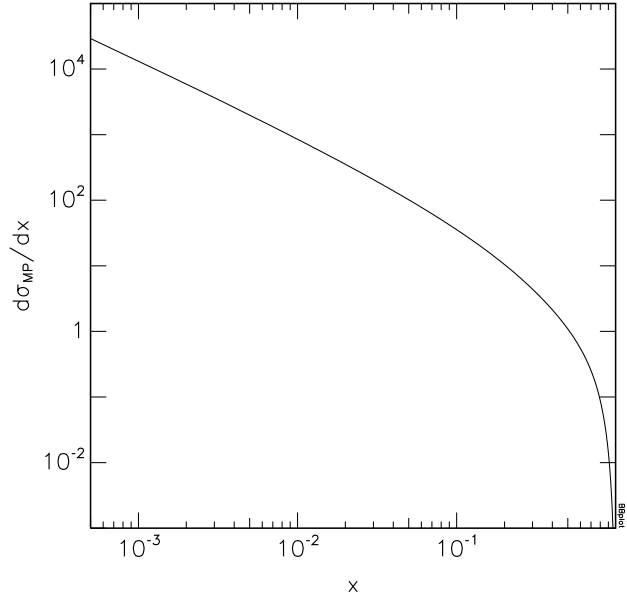,height=80mm}
  \caption{Ortho-Ps spectrum in electroproduction in units
  of $(4 \alpha /\pi) \sigma_0 \, \zeta (3) \, \nu^2\, B(\nu)$ due to
   multi-photon mechanism of Fig.~\ref{Fig:2}.}
  \label{fig7}
\end{figure}

\begin{figure}[!htb]
  \centering
  \epsfig{file=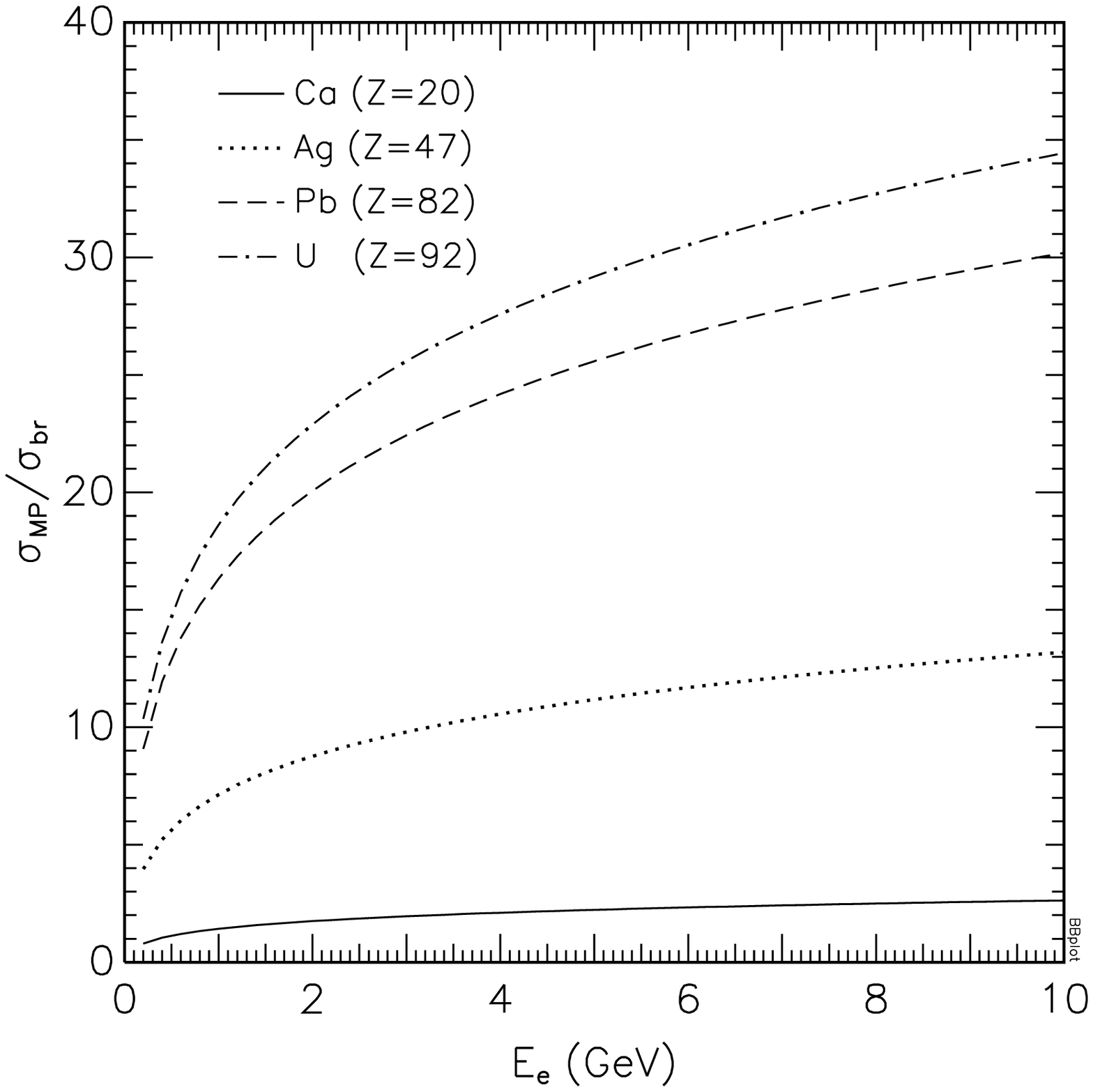,height=80mm}
  \caption{Ratio $\sigma_{\mathrm {MP}}/ \sigma_{\mathrm {br}}$ of
  multi-photon (Fig.~\ref{Fig:2}) to  bremsstrahlung
cross section  (Fig.~\ref{Fig:3}) as function of the electron beam energy
$E_e$.}
\label{fig8}
\end{figure}

At the end we would like to note that our results cannot be straightforwardly 
transformed to the production of $\mu^+ \mu^-$ elementary atom called dimuonium
(DM). For the photo- and electroproduction of DM a new
important phenomenon takes place, namely, the restriction of the
transverse momenta $k_{1\perp},\dots, k_{j\perp}$ for the exchanged
photons in Figs.~\ref{Fig:1}, \ref{Fig:2}. This restriction arises
due to the nucleus form factor at the level
$\stackrel{<}{\sim}  1/r_A \ll m_\mu$  where $r_A$ is the
electromagnetic radius of the nucleus. As a result, the effective
parameter of the perturbation theory becomes small $\sim
\nu^2/(r_A m_\mu)^2 \stackrel{<}{\sim} 0.03$, contrary to the Ps
case.
A detailed study of DM production will be presented in \cite{GJKKSS}.

\acknowledgments

We are grateful to R.~Faustov, I.~Ginzburg, I.~Khriplovich,
I.~Meshkov and L.~Nemenov for useful discussions and to
A.~Arbuzov, O.~Krehl and B.~Shaikhutdenov for help. The work of
S.R.G. and E.A.K.  was supported by INTAS grant 93-239 ext, the
work of V.G.S is supported by Volkswagen Stiftung (Az. No.1/72
302) and by Russian Foundation for Basic Research (grant
96-02-19114).

\begin{thebibliography}{99}

\bibitem{M}
I. N. Meshkov, Elem. Particles and Nuclei {\bf 28}, 495 (1997).

\bibitem{KM}
I. B. Khriplovich, A. I. Milstein, Novosibirsk preprint Budker-INP 96-49
(1996).

\bibitem{N}
L. L. Nemenov, Yad. Fiz. {\bf 51}, 444 (1990), 
[Sov. J. Nucl. Phys. {\bf 51} 284 (1990)]; 
V. L. Lyuboshitz, M. I. Podgoretsky, ZhETF {\bf 81},
1556 (1981).

\bibitem{MSS}
G. V. Meledin, V. G. Serbo, A. K. Slivkov, Pis'ma ZhETF {\bf 13}, 98
(1971) [JETP Lett. {\bf 13}, 68 (1971)].

\bibitem{O}
H. A. Olsen, Phys. Rev. {\bf D33}, 2033 (1986).

\bibitem{TC}
A. V. Tarasov, I. V. Christova, Dubna preprint JINR P2-91-4, (1991).

\bibitem{AB}
A. A. Akhundov, D. Yu. Bardin, L. L. Nemenov, Yad.Fiz. {\bf
27}, 1542 (1978).

\bibitem{HO}
E. Holvik, H. A. Olsen, Phys. Rev. {\bf D35}, 2124 (1987).

\bibitem{IM}
D. Ivanov, K. Melnikov, Phys. Rev. {\bf D57}, 4025 (1998).

\bibitem{BMOM} 
H. A. Bethe and L. C. Maximon, Phys. Rev. {\bf 93}, 788 (1954);
H. A. Olsen and L. C. Maximon, Phys. Rev. {\bf 114}, 887 (1959).

\bibitem{BLP}
V. B. Berestetskii, E. M. Lifshitz, L. B. Pitaevskii, {\it Quantum
Electrodynamics} (Nauka, Moscow, 1989).

\bibitem{NOSVVZ}
V. A. Novikov et al., Phys. Rep. {\bf 41C}, 1 (1978).

\bibitem{BGMS}
V. M. Budnev, I. F. Ginzburg, G. V. Meledin, V. G. Serbo, 
Phys. Rep. {\bf C15}, 181 (1975).

\bibitem{GJKKSS}
I. F. Ginzburg, V. D. Jentschura, S. G. Karshenboim, F. Krauss, V. G. Serbo, 
G. Soff, hep-ph/9805375, Phys. Rev. {\bf C} (in print).

\end{thebibliography}
\end{document}